\documentstyle [12pt]{article}
\evensidemargin\oddsidemargin
\begin{document}
\count0 = 1
\begin{titlepage}
\vspace{30mm}
 \title{\small{  QUANTUM MEASUREMENTS, INFORMATION 
AND DUAL STATES REPRESENTATIONS \\ }}
\author{S.N.Mayburov 
 \thanks{E-mail ~~ maybur@sgi.lpi.msk.su  ~~
 }\\
Lebedev Inst. of Physics\\
Leninsky Prospect 53\\
Moscow, Russia, 117924\\
\\}
\date {22.09.2000}
\maketitle
\begin{abstract}
The  quantum measurement problem considered
for the model of measuring system (MS) 
 consist of measured state S (particle),
 detector D and information processing device (observer) $O$
interacting with S,D. 
 For 'external' observer $O'$  MS evolution
  obeys to Schrodinger equation (SE) and $O$ (self)description
 of MS reconstructed from it in Breuer ansatz. 
 MS irreversible evolution (state collapse) for $O$ 
 can be obtained  if the true  quantum states manifold
has the dual structure $L_T=\cal {H} \bigotimes L_V$
where $\cal H$ is  Hilbert space and $\cal L_V$
is the set with elements
$V^O=|O_j\rangle \langle O_j|$ describing random 'pointer'
outcomes $O_j$  observed by $O$ in the individual events.
Possible experimental tests of this dual states structure described.
 The results interpretation in Quantum Information framework and
 Relational Quantum Mechanics discussed.
\end{abstract}
\vspace{20mm}
\small{Talk given on 'Quantum communications and measurements' conference\\
Capri ,  July 2000, (Kluwer, N-H,2001)\\}
\vspace {20mm}
%\small{ Int. Journ. Theor. Phys. to appear (1999)}
\vspace{20mm}
\end{titlepage}
\section { Introduction}
Despite that Quantum Mechanics (QM) describe perfectly most of
experimental effects in microscopic domain there still some
difficult questions and 'dark spots' connected with its  foundations
and in particular Quantum Measurement Theory.
 Of them the problem of  the state vector collapse or the objectification
in general seems  most remarkable  and it's
still unresolved despite the multitude of the proposed
solutions ( for the review see $\cite  {Busch}$). Eventually
the measurements and collapse studies can help us to
 select the true QM interpretation out of many proposed.
 This paper analyses the information transfer  to observer
in the measurement and the collapse theory which is prompted by this
considerations. We regard the microscopic dynamical model of 
measurements in which 
  the  evolution of the measuring system (MS) described from
the first QM principles. In our approach  MS in general includes
the measured state (particle) S,
  detector D  amplifying S signal  and
 observer $O$ which processes and stores the information. Under 
observer we mean information gaining and utilizing system 
(IGUS) of arbitrary structure  $\cite {Gui}$.
 It can be  human brain or some automatic device 
processing the information , but in both cases  practically it's
the system with many internal degrees of freedom (DF) on which
 the large amount of information can be memorized.
In general the computer information processing or perception  by human brain
 is the physical objects evolution which on microscopic level
supposedly obeys to  QM laws   $\cite {Alb}$.

Standard Copenhagen QM interpretation divide our physical world
into microscopic objects which obeys to QM laws and macroscopic objects
, including observers which supposedly are  classical.
 This artificial partition
was much criticized, first of all because it's not clear where
to put this quantum/classical border. Moreover there are strong
experimental evidences that at the dynamical level no such border
exists and QM  successfully describes large, complicated systems
including biological ones.    

The possible role of observer  
in quantum measurements was discussed for long time \cite {Wig}, but 
now it attracts the significant attention also  due to the
the progress of quantum information studies $\cite {Pen}$. 
The class of microscopic measurement theories which attempts
to describe observer quantum effects sometimes called Relational QM
( for the review see  $\cite {Rov}$).
% Witnessing QM interpretation \cite {Koch}.
In general Relational QM  concedes
 that QM description is applicable
both for microscopic  states  and macroscopic
objects including observer $O$ (he, Bob) which  Dirack state vector
$|O\rangle$ ( or density matrix $\rho$ for more realistic cases)
 can be defined relative to  some other observer $O'$ (she; Alice), 
which is also another quantum object.
Of course this assumption it's not well founded and real $O$ state
can be much more complicated, but it's reasonable to start from
 that simple case.
 Consequently the evolution of any complex system C
 described by Schrodinger equation 
  and for any  C
 the superposition principle holds true 
at any time.
 MS measurement description by $O$ 
formally  must include evolution of 
 $O$  own internal DFs  which  participate in the interaction with S 
 $\cite {Alb}$. 

The role of observer in the measurement and its selfdescription
called also selfmeasurement
often  regarded as the implication
of the general algebraic and logical problems of selfreference $\cite {Busch}$.
Following this approach Breuer derived the general selfmeasurement 
restrictions  for classical and quantum measurements  $\cite {Bre}$.
This formalism don't resolve 
the measurement problem completely,  but 
   results in some important restrictions on its possible solution.
 From this analysis we propose modification of standard
QM Hilbert space formalism  which aim is to describe measurement process
consistently.
Its  main feature is the extension of QM states manifold permitting
to account observer selfmeasurement features.  
Of course if some correction to quantum dynamics like in GRW model
exist $\cite {Gir}$ then the state collapse can occur in macroscopic
detectors. But until such effects would be found and the 
standard QM Hamiltonian can be regarded well established,
 we must inspect
in detail  observer properties in search of collapse.

In chap. 2 we describe our measurement model in detail.
We argue that collapse description demands
to change standard  QM formalism
and propose the particular variant of its modification.
In chap. 3 we'll discuss $gedankenexperiments$ which
help to interpret our formalism and discuss its meaning.
 In chap. 4 
physical and philosophical implications
of this results and interpretation discussed.

Here it's necessary to make some technical comments on our model premises
and review some terminology.
In our  model we'll suppose that MS  always can
 be described completely (including Environment E if necessary)
 by some state vector $|MS\rangle$ relative to 
$O'$ or by density matrix if it's in the mixed state.
 MS can be closed system , like atom in the box or open pure
system  surrounded by electromagnetic vacuum or  E of other kind.
We don't assume in our work any special dynamical
 properties of $O$ internal states beyond standard QM. 

  In this paper  the brain-computer analogy used without
discussing its reliability and philosophical 
implications $\cite {Pen}$. We'll ignore here quantum computer options
 having in mind only the standard  dissipative computers.
We must stress that throughout our paper the observer consciousness
 never referred directly.
 Rather in our model observer can be regarded as active 
reference frame (RF) which interacts with studied object.
Thus S state description 'from the point of view' of the particular $O$
described by the terms 'S in $O$ RF' or simply 'for $O$'.
 The terms 'perceptions', 'impressions' used by us
 in a Wigner sense $\cite {Wig}$ of observer subjective description
of experimental  results 
and  can  be defined in strictly physical and Information
theory terms ( more discussion given in Conclusion).

\section {Selfmeasurement and Weak Collapse}
 
 To perform the measurement one needs detector D and IGUS $O$ 
each of them in practice have many internal DFs.
In elementary measurement model  the detector D and observer
 $O$ only with one  DF 
regarded  corresponding to Von Neuman scheme.
Account of many DFs for D and $O$ doesn't change principally the results
obtained below $\cite {May3}$, but
in addition it resolves the problem of 'preferred basis'
arising for one  DF detector model $\cite {Elb}$.  
The example of dynamical model with many DFs
gives Coleman-Hepp model described in  $\cite {Hep}$.

 Let's consider in this one DF ansatz  $O'$ description of
 the measurement performed by $O$ of binary observable $\hat{Q}$ on S state : 
 $$
\psi_s=a_1|s_1\rangle+a_2|s_2\rangle
$$
, where $|s_{1,2}\rangle$ are $Q$ eigenstates with values $q_{1,2}$.
 For the  simplicity in the following  we'll  omit detector D in 
 MS chain
assuming that S directly interacts with $O$. It's possible for our 
simple model
because if to neglect decoherence the only  D effect is
the amplification of S signal to make it conceivable for O.
Initial  $O$ state  is $|O_0\rangle$
relative to $O'$  RF 
and  MS initial state is :
\begin {equation} 
     \Psi^0_{MS}=(a_1|s_1\rangle+a_2|s_2\rangle)|O_o\rangle  \label {AAB}
\end {equation}
We assume that  S-$O$ measuring interaction starts at $t=t_0$
and finished at some finite $t_1$.
 From the linearity of Schrodinger equation
 for suitable interaction Hamiltonian $\hat{H}_I$
 for  some $t_1$ at $t>t_1$ 
 the state of MS system  relative to $O'$ observer will be $\cite {Busch}$
\begin {equation}
   \Psi_{MS}=a_1|s^f_1\rangle|O_1\rangle+
a_2|s^f_2\rangle |O_2\rangle
                                   \label {AA2}
\end {equation}
to which corresponds the density matrix $\hat{\rho}_{MS}$.
Here  $|O_{1,2}\rangle$ are $O$ state vectors
obtained after  the measurement of particular $Q$
eigenstate  $|s_{1,2}\rangle$  and are the eigenstates
of $ Q_O$ 'pointer' observable.
In most cases one can take for the simplicity $|s^f_i\rangle=|s_i\rangle$
without influencing main results.

 All this states including $|O_i\rangle$ belongs to Hilbert space
$\cal H'$ defined in $O'$ RF and Hilbert space $\cal H$ in $O$ RF 
can be   obtained performing unitary $\cal H'$ transformation $\hat{U}'$
 to $O$ c.m.s.( below state vectors with $n>2$ components used with the
same notations). $\hat{U}'$  can be neglected if only internal
or RF independent discrete states regarded permitting to take $\cal H=H'$.
We suppose that for $t>t_1$ measurement definitely finished which
simplifies all the calculations
, but in fact that's fulfilled exactly only for the restricted
class of models like Coleman-Hepp. 

Thus QM predicts  at time $t>t_1$  for external observer $O'$ 
MS is  in the pure state  $\Psi_{MS}$ of (\ref {AA2}) which is superposition of
 two states. 
Yet we know from the experiment that 
macroscopic   $O$ observes some definite random $Q_O$ value $q^O_{1,2}$
 from which he concludes
that  S final state is $|s_1\rangle$ or $|s_2\rangle$.
In standard QM  with Reduction Postulate MS final  state coincides with
the statistical ensemble of such individual final states for $O$
described by density matrix 
 of mixed state $\rho^s_m$ which  presumably
means the state collapse:

\begin {equation}
 \rho^s_m= \sum_i |a_i|^2|s_i \rangle \langle s_i|
                                                              \label {AA33}
\end {equation}
In our ansatz where where $O$ regarded as quantum object interacting with S 
we can ascribe to MS the corresponding mixed state :
\begin {equation}
 \rho_m= \sum_i |a_i|^2|s_i \rangle \langle s_i||O_i \rangle \langle O_i|
                                                              \label {AA3}
\end {equation}

Normally the states in two RFs are connected by some unitary transformation
, but
no such transformation between (\ref {AA3}) in $O$ RF
and (\ref{AA2}) in $O'$ RF is possible $\cite {Busch}$.
It's quite difficult to doubt both in correctness of
 $O'$ description of MS evolution
by Schrodinger equation and in collapse experimental observations.
If observers regarded as quantum objects and accounted in measurement chain
 then this  contradiction constitutes famous Wigner dilemma
 or 'Friend  Paradox' for $O, O'$  $\cite {Wig}$. From $O$ 
'point of view' $\Psi_{MS}$ describes superposition of two
contradictory impressions : $Q=q_1$ or $Q=q_2$ percepted simultaneously. 
 This paradox prompts to investigate 
possible QM formalism modifications and first one should investigate
QM formalism of system description 'from inside'.

 For realistic IGUS $|O_{1,2}\rangle$ can correspond to some
excitations of $O$ internal collective DFs
 like phonons, etc., which memorize this $Q$ information,
 but we don't consider its particular physical mechanism here. 
It's necessary to formulate in our theory
some minimal assumptions about relation between the observer  states  and
 his  perception  of the input quantum signal.
We'll suppose that for any Q eigenstate $|s_i\rangle$ 
 after S measurement finished at $t>t_1$ 
and $O$ state is $|O_i\rangle$ observer $O$ have the impression
that  the measurement event occurred and the value of
outcome is $q_i$.   
  If S state is the superposition $\psi_s$ then we'll
suppose that its measurement also  result in appearance of some arbitrary $O$ 
impression which properties will be discussed below.
We stress that we don't suppose any special properties of 
biological systems. The simplest $O$
toy-model of information memorization is hydrogen-like atom
 for which $O_0$ is ground state
and $O_i$ are the metastable  levels excited by $s_i$, resulting into
final S - $O$ entangled state.
This considerations have little importance for the following formalism
rather they explain our philosophy of impressions-states relation.

Here we assumed that at $t>t_1$ measurement finished with probability
$1$, but for realistic measurement Hamiltonians
 it's only approximately true, because  transition amplitudes have 
long tails. This assumption exactly fulfilled only for some simple
models like Coleman-Hepp, but we'll apply it here  to simplify our analysis.   
The more subtle question of exact time at which $O$ percepts its
own final state $O_j$ isn't important at this stage.
We'll assume  that after S-$O$ interaction S leaves $O$ volume, which
can be regarded as MS 'self-decoherence'. Thus  final MS state quantum
phase becomes unavailable for $O$. The practical decoherence
mechanisms and their effects will be
 discussed in final chapter $\cite {May3}$.

To discuss $O$ selfdescription  let's start with Breuer selfmeasurement
theorem which is valid  both for
classical and quantum measurements $\cite {Bre}$.
Any measurement of studied system MS is the 
mapping of MS states set $N_S$ on  observer states set $N_O$.  
For the situations when observer $O$ is the part of the studied system MS 
- measurement from inside, $N_O$ is $N_S$ subset and 
$O$ state in this case is MS state projection on $N_O$
 called MS restricted  state $R_O$.  
From $N_{S}$ mapping properties the principal restrictions for
$O$ states were obtained in Breuer theorem. It proves 
  that if for two arbitrary MS 
states $S_{MS}, S'_{MS}$ 
their restricted  states $R_O, R'_O$ coincide then for $O$ this MS 
states are indistinguishable.
 The origin of this effect is easy to 
understand : $O$ has less number of DFs then MS and so can't describe
completely MS state.
For quantum measurements $ O$  restricted state
can be the partial trace of complete MS state  (\ref {AA2}) :
\begin {equation} 
   R_O=Tr_s  \hat{\rho}_{MS}=\sum |a_i|^2|O_i\rangle\langle O_i|
      \label {AA4}
\end {equation}
$R_O$ can be interpreted as $O$ subjective state which
describe his subjective perception.
Note that  for MS  mixed state $\rho_{m}$
of (\ref {AA3}) the  corresponding restricted state is the same $R^m_O=R_O$.
This equality doesn't mean collapse of MS state $\Psi_{MS}$, because
it holds for statistics of quantum ensembles, but collapse appearance
  also must be  tested specially  for individual events states.
Such restricted $R_O$ form assumes
 that O can percept only his internal
excitations independently of quantum correlations with  S state.
This assumption can be wrong for quantum
systems due to well known quantum entanglement
and in fact  its study shows  that from equality of 
restricted states doesn't follows the transition of pure system 
 state to mixed one.

Exploring individual events it's important to note that for mixed incoming
 state S MS individual state   in event $n$  differs from
statistical state (\ref{AA3}) and is equal to :
$$
\rho^m_n=|O_l\rangle \langle O_l|| s_l\rangle\langle s_l|
$$
with arbitrary, random $l(n)$ which in  standard QM objectively
exists in event $n$, but can be initially unknown for $O$  $\cite {Bre}$.
Its restricted state is $R^n_O=|O_l\rangle \langle O_l|$
 and so differs always from 
$R_O$ of (\ref {AA4}). Due to it main condition of Breuer  Theorem violated
and so this theorem isn't applicable for this situation
and $O$ can differentiate pure/mixed states 'from inside'.
 It means that the restricted states
ansatz doesn't prove the collapse appearance in standard QM
 formalism  even with inclusion of observer quantum effects in the
measurement models. The analogous conclusions
 follows from the critical analysis of Witnessing
 interpretation $\cite {Koch,Lah}$. As we noticed already statistical
restricted states for mixed and pure states $R_O^m,R_O$ coincide,
but it doesn't mean as Breuer claims that in ensemble observer $O$
can't differ pure and mixed state. The reason is that $O$ can
analyse ensemble properties  not only statistically,
 but on event by event basis 
regarding $N$-events restricted state which is tensor product of $N$
 $R^n_O$ states
which all components differ from $R_O$. Really it would be strange
otherwise if in each event pure and meixed state  differs,
 but ensemble of events don't reveal any difference.

Additional arguments in favor of this conclusion
reveals  MS interference term observable :
\begin {equation}
   B=|O_1\rangle \langle O_2||s_1\rangle \langle s_2|+j.c.
    \label {AA5}
\end {equation}
being measured by $O'$ gives $\bar{B}=0$ for mixed MS state (\ref {AA3})
, but in general $\bar{B}\neq 0$  for pure MS state (\ref{AA2}).
It  evidences that even for statistical ensemble 
the observed by $O'$ effects  differentiate  pure and mixed MS states.
Note that $B$ value principally can't be measured by $O$ directly, because
$O$ performs $Q$ measurement and $[Q,B] \neq0$ $\cite {May3}$.

 Breuer analysis is quite  useful, because  in fact
it prompts the minimal modification of MS states set $N_S$
which can describe MS state collapse. Here we'll demand that
 QM modification in our theory satisfy to two main operational conditions : \\
i) if S (or any other system)  don't interact with $O$
 then for $O$ this  system evolves according to Schrodinger equation
dynamics  (SD) (for example GRW theory  brokes this
condition \cite {Gir}).\\
ii) If S interacts with $O$ (measurement)  SD can be violated for $O$
so that he percepts random events, but in the same time
 as $O'$ doesn't interact with S or $O$
 i) condition settles  that in $O'$ RF  MS must be described by SD.
 It means that in such formalism MS  final states
 relative to $O$ and $O'$ are nonequivalent $\cite {Busch}$.
  We'll call this phenomena  the weak (subjective)
collapse  which obviously has  looser conditions then standard QM 
Reduction Postulate.
We attempt to satisfy to both this conditions performing minimal
 modification of QM states set - which  in standard QM is
Hilbert space.
This modified set must incorporate simultaneously both linear states
evolution and random events observed by $O$, but Schrodinger
dynamics must be conserved copiously if interaction with observer
absent. It's worth to remind that Hilbert space
is in fact empirical construction which choice advocated by fitting
most of QM data, and so QM states set modification isn't
 unthinkable in principle. Such attempts were 
published already and most famous is Namiki-Pascazio many Hilbert spaces
 formalism \cite {Nam}.
 In standard QM formalism all its states manifold representations
are unitarily equivalent, but observers interactions and
evolution aren't considered in it.
It will be shown that our new formalism 
  in some sense is analog of the
nonequivalent representations of states manifold. Analogous
superselection systems
are well studied for nonperturbative Field theory (QFT)
 with infinite DF number
$\cite {Ume}$. This approach was applied for measurement problem,
 but it's not clear its applicability for finite 
systems $\cite {Fuk,May3}$.

Remind that  experimentalist
never observes state vector directly, but his data consists of
individual random events like detector counts and 
  the state vector reconstructed  from
observed random events statistics. Following this notion
we'll suppose  that the  perception by $O$  
of individual events  to large extent is independent of system state vector
 $\Psi_{MS}(t)$.
 This  prompts to regard 
 QM dual representations, in which state vector and observer information
presented simultaneously by  independent entities. 
To illustrate the formalism features and introduce its terminology
let's describe as example how such formalism describes
 $O$ measurement of some  parameter $q$ with probabilistic
 distribution $P^c(q)$ which describe the classical statistical ensemble.
For such $P^c(q)$ distribution  (or $P^c_l$  array for discrete $q$ )
when $O$ measures $q$ he acquires instantly information about $q$ value
and initial $O_0$ state changes to some $O_j$ correlated with
measured $q_j$. Formally
 at this moment $P^c(q)$ collapses to delta-function, but
    in classical case
 this effect reflects only $O$ information change.
For $q$  discrete 
 the observer information  $O(n)=O_i$ in given event $n$ presented by
  1-dimensional matrix  $V_l^O=\delta_{li} O_i$.   
For  complete system description we  can use formally the dual 
 event-state  
 as $\Phi_{clas}=P^c \bigotimes V^O$.   
  It  incorporates statistical state of ensemble
 $P^c$ and $O$ information in the individual event $n$.     
and such dual form acquires nontrivial meaning
for our modified QM states set.
Note that $V^O$ by itself is unavailable to other $O'$, which can get
this information only via signalling between this two observers, i.e.
 their interaction whcih in principle must be accounted.

  To explain the main idea for the beginning we regard   this
 new formalism  applied to our MS system evolution.
 $O$ and $O'$  Hilbert spaces $\cal H, H'$ will be our initial basis,
and  QM density matrices manifold $L_q= ( \rho \geq 0, Tr \rho=1)$
constructed of $\cal H$ state vectors will be used.
Analogously to classical example $\rho$ is one component
 of our dual state $\Phi(n)= \rho \bigotimes V^O(n)$ and $V^O$ describes the 
 outcome of individual event. Alike in standard QM $\rho$
 obeys  always including measurement process
to  Schrodinger-Liouville equation  
 for arbitrary MS Hamiltonian $\hat{H}_c$ :
\begin {equation}
    \dot\rho=[\rho,\hat {H}_c]  \label {AA8}
\end {equation}
which for pure states is equivalent to Schrodinger equation.
Initial $\rho(t_0)$ states defined also by standard QM rules.
 Inside $L_q$ we extract $O$ restricted
states  analogously to (\ref {AA4}) $R_O=Tr_s \rho$ and calculate in $O$ basis
weights matrix 
\begin {equation}
P_j(t)=Tr (\hat{P}^O_j R_O)=Tr (\hat{P}^O_j \rho(t))  \label {BB99}
\end {equation}
 where $\hat{P}^O_j$
is $O_j$ projection operator. For (\ref {AA2})
and $t>t_1$ it  gives $P_j=|a_j|^2$ equal to standard QM 
reduction probabilities.
We suppose that   $P_j$ is 
the probabilistic distribution  which describes individual events outcomes
$O_j$ percepted  by  $O$  after S-$O$ interaction. In distinction from
standard QM 
this subjective $O$ information  in the individual event is only statistically
correlated with final $\rho(t)$ at $t>t_1$ and 
described by random vector
 $V^O=  |O_j\rangle \langle O_j|$ in the individual event 
% $V^{Oj}=|O_j\rangle \langle O_j|$
 corresponding to observation  of  random $Q$ value $q_j$.
Despite that $O$ percepts some random outcome $O_j$
 due to regarded $V^0$ independence  $\rho (t)$ 
 evolves all the time according to Schrodinger-Liouville equation
 and  doesn't suffers the abrupt state collapse to $\psi_j$.
Thus our dual state or event-state, which due to $j(n)$ randomness
 differs for each individual event $n$ for $O$  is doublet of
dynamical and information component :
\begin {equation} 
   |\Phi_n\gg=|\phi_D, \phi_I \gg=
|\rho, |O_j\rangle \langle O_j|  \gg \label {A22}
\end {equation}
 $V^O$ has the special 'information'
dynamics regarded below and only statistically correlated with $\phi_D$.
 Before measurement starts
 $O$ state  is $|O_0\rangle$ and the  dual state is
 $\Phi=|\rho_{MS}(t_0), V_0^O \gg $ where
 $V_0^O=|O_0\rangle \langle O_0|$, so that
initial $O$ information  described both by
 $V_0^O$ and $\rho_{MS}(t_0)=|\Psi^0_{MS}\rangle \langle \Psi_{MS}^0|$
The  time of $V^O_0\rightarrow V^O_j$ transition for $O$ is between
$t_0$ and $t_1$  and can't be defined in the current  formalism with larger
accuracy, but it doesn't seems very imortant at this stage.
  $O'$ doesn't interact with MS and due to it  MS final state for her is
$\Psi_{MS}$ of (\ref {AA2}) as regarded in detail below. 

 Probabilities $P_j$ coincides with corresponding standard QM 
probabilities  $P^Q_j=Tr (\hat{P}_{Sj}|\psi_s\rangle \langle \psi_s|)$
of particular outcome $q_j$.
\cite {Busch}. 
If we restricts  only to statistical ensemble description
and aren't interested in particular event outcome
then statistical dual state can be defined :
$$
|\Theta^s)=\rho \bigotimes\ \{V^P \}
$$
where $\{V^P\}$ is $V^P_j=P_j$ vector, describing probabilistic
 distribution of$V^O= |O_j\rangle \langle O_j |$ outcomes in $L_V$ subset.

Complete manifold in $O$ RF for this event-states is
$N_T=L_q \bigotimes L_ V$ i.e direct product of dynamical and
subjective components. $L_V$ is the  linear space
of diagonal positive matrices or vectors $V^O$ with $tr V^O=1$
and consequently $|\Phi|=1$ for all such states.
 If we restrict our consideration only to
pure states as we do below then $N_T$ is equivalent
  to $\cal {H} \bigotimes \mit L_V$
 and the state vector can be used as the  dynamical component $\phi_D$.
 $O_j$    entangled with $s_j$ and 
 in place of $V^O$  equivalent to it MS   subjective dual state
component
  $V^{MS}=  |O_j\rangle \langle O_j| |s_j\rangle \langle s_j|$
can be used.

Naturally in this formalism
$O'$ has her own subjective space $L'_V$   and in her RF  the events states
 manifold is $N'_T=\cal H' \bigotimes \mit L'_V$ for pure states.
 From the described features it's clear that subspace
$L_V$ is principally unobservable for $O'$ (and vice versa for $L'_V,O$),
 because in this formalism only  the measurement
of $\phi_D$ component described by eq. (\ref {AA8}) 
permitted for $O'$. 
But $V^O,V'^O$ can be correlated statistically via special measurement by $O'$
 of dynamical component $\phi_D$. For this purpose $O'$ can measure
$Q_O$ on $O$ getting the information on $V^O$ content,
  which mechanism we'll discuss in next chapter.
Formally $O'$ also can ascribe to MS $V^O$ internal coordinate
but $j$ value is uncertain for her in $\Psi_{MS}$ final state and so
it has little sense.
It means that $O'$ is sure that $O$ knows $q_j$ value,
but $O'$ don't know this value.   
 In general if in the Universe
altogether N observers exists then the complete states manifold
described in $O$ RF  is 
$L_T=\cal {H} \bigotimes \mit L_V \bigotimes L'_V ...\bigotimes L^N_V$
of which only first two subsets are observed by $O$ directly and
all others available only indirectly via $\cal H$ substates.

Standard QM
reduction postulate also describes  how the state vector
correlates with the changes of observer information in the measurement.
The main difference is  that in place of abrupt and irreversible
 state vector $\psi_s$ reduction to some random state vector $\psi_j$ in 
standard QM in our formalism  the dynamical component $\Psi_{MS}$
of  MS event-state evolves linearly and reversibly
in accordance with (\ref {AA8}). 
It's only subjective component which changes abruptly and 
probabilistically describing $O$ subjective information about $S$.
We must stress that 
subjective or informational $V^O$ component of $\Phi_n$ 
 isn't the new degree of freedom, but   $O$ internal coordinate $Q_O$
 which (self)description relative to $O$
 doesn't covered by MS Hilbert space $\cal H$  vector $\Psi$ only.
 $Q$ information  for $O'$ which don't interact
with S completely described by $\Psi_{MS}$, which means uncertainty
$q_{min} < Q < q_{max}$.  $V^O$  substate contains additional information
$Q=q_j$ available for $O$ only.
In such theory $O$ can percept only (random) part of the full
state vector $\Psi_{MS}$ which continue to exist.
 Schrodinger dynamics and collapse coexist, by the price
 that S signal perception by $O$ occurs via this new stochastic
   mechanism. Despite we use term 'perception' in our model it doesn't 
referred to human brain specifically. We believe that as $O$ can be
regarded any system which  can produce the stable entanglement of
its internal state and measured state S. As we mentioned already
it can be even  hydrogen-like atom in the simplest model for 
which $O_i$ can be different atomic levels.  Perception
corresponds to $V^O$ component presence which means that in such formalism
 $O$ states completely
described only by $O$ 'from inside'  and  not by any other $O'$.
 
Now let's describe  dual  formalism for pure states in details.
 If S don't interact with $O$ (no measurement)
 then $V^O$ is time invariant
 and one obtains
standard QM evolution for event-state dynamical component $\phi_D=\Psi$
-  state vector.
Thus our dual states are important only for measurement-like  processes
with direct $O$ interactions, but in such case it's always  the 
 analog of regarded MS system. Its dual state is
$|\Phi \gg =|\Psi_{MS} ,V^O\gg$ and its evolution defined by initial
$\Psi_{MS}(t_0)$ and the formal system of equations :

\begin {eqnarray}
i\dot{\Psi}_{MS}=\hat{H}_c \Psi_{MS}    \nonumber \\  
P_j=Tr( \hat{P}^O_j |\Psi_{MS} \rangle \langle \Psi_{MS}|)  \label {CC} \\
l=Rnd(P) \nonumber  \\  
V^O=|O_l \rangle \langle O_l| \nonumber
\end {eqnarray}
where $Rnd(P)$ is the random numbers generator for $P$ distribution
producing random index $l(n)$ in the individual event $n$.
\cite {Jan}. The statistical dual state $|\Theta^s)$ evolution
 defined by first two equations. The first
 equation of (\ref {CC}) is Schrodinger equation which becomes here
the analog of master equation for
probabilities $P_j$ describing $V^O$ probabilistic  distribution.

Due to independence of MS dynamical state component $\phi_D$
of internal parameter $j$ of $V^O$ 
this $O$-S $\Phi$ evolution is reversible. Thus in dual formalism
 no experiment performed
by $O'$ on MS wouldn't contradict to standard SD. If $O$ perform
selfmeasurement experiment on MS the situation is more subtle and
 will be discussed in the next chapter.  Note that in this formalism
parameter $j$ don't existed before S-$O$ interaction starts.

Above $L_V$ corresponds to the simplest measurement and in general it  
will can have much more complicated form corresponding
to $O$ information channels and $O$ structure. As we noticed their parameters
must correlate to
the complex  probabilities of standard QM $\cite {Busch}$. For
example 2-dimensional values correlation measurement by $O$
 has the distribution :
\begin {equation}
P_{ij}=Tr(\rho\hat{P}^O_{1i}\hat{P}^O_{2j}) \label {AA22}
\end {equation}
where $\hat{P}^O$ are projectors on $O$ memory states
$|O^1_i\rangle, |O^2_j\rangle$.
Corresponding set structure is $L_V=V^{O1}\bigotimes V^{O2}$.
  
Extension of dual formalism on completely mixed states is obvious and  
here it presented  only for the states interested for  measurements of the kind
(\ref {AA3}). For them from eq. (\ref {BB99}) naturally follows 
$P_j=|a_j|^2$ which gives $V^O$ distribution.
In dual formalism the restricted MS state
 $R^V=|O_l\rangle  \langle O_l|$  where 
$l(n_1)$  in  the individual event $n_1$ defined by (\ref{CC}). It
differs from restricted state $R_O$ in standard QM given by (\ref {AA4}),
but coincide with restriction of mixed state $\rho^m_n$
 in the individual event $n$ if $l(n_1)=l(n)$. Thus Breuer
theorem condition fulfilled in dual formalism as expected.

Obviously in this formalism $\bar{Q}$ coincide both for $O$ and $O'$.  
If one interested only to calculate $\bar Q $ after S measurement by $O$ 
or any other expectation values ignoring event structure
it's possible to drop $V^O$ component and to make standard QM calculations
for $\rho$.
If one regards the statistical results for quantum ensembles
then statistics in $L_V$ subspace corresponds
to  $|a_j|^2$ the probabilities of particular $O$
observation. Note  that their meaning differs from $O'$ 
representation where they can't be regarded as probabilities
but only like some weights. Because $O$ 
observes random $Q$ values then if 
 in addition to demand that $\bar{Q}$ coincide both for $O$ and $O'$
then one obtains that $Q$ distribution for $O$ described by $|a_j|^2$.
Thus dual formalism gives naturally
the values of outcome  probabilities $|a_i|^2$ which is quite difficult
to obtain in some theories explaining state collapse like
Many Worlds Interpretations (MWI) $\cite {Busch}$. 

 To exclude spontaneous $V^O$ jumps
without effective interactions with external world 
 we introduce additional $O$ identity condition :
 if S and O don't interact   then
 the same  random  parameter j of $V^O$ conserved.
We extend it to more general condition : if  different $\Psi_{MS}$
$O_j$ branches don't intersects i.e.
 $\langle O_i|\hat{H}|O_j\rangle=0 , i \neq j$ then $V_J^O$ conserved.  
As we noted it means that $O$ observes 
 constantly only $|s_j\rangle $ branch of S state.
 Note that this
condition doesn't influence on MS dynamics defined by $\rho$, but
only on dynamics of $O$ information $V^O$.
Such condition is natural if $P_i$ can be regarded as transition
probabilities $V^O_0 \rightarrow V^O_i$ resulting from action of $\hat{H}_I$.
   Really $P_i$ can be rewrited as :
$$
   P_i=|\langle \Psi_i|U(t_1-t_0)|\Psi^0_{MS}\rangle|^2
$$
where $\hat{U}$ unitary trasformation in time. In this case if S-$O$
stops to interact $\hat{H}_I(t)=0$ then all $P_{ij}=0$ and it's 
compatible with absense of such jumps. 

Note that in general the theory can be formulated even without
this condition, because $O$ or any other observer
 can't indicate that such jumps occurs.
The reason is that in our model $O$ memory described by $V^O$ and
if the jump $j \rightarrow i$ occurs no memory about previous state 
can be conserved. For more complicated memory structure such
jumps should occurs simulteneosly in many cells with entangled states. 
Thus this  condition simplifies the theory and make it more reasonable
but in general isn't necessary.

 In our formalism parameter $j$ of $V^O$
defined at random with probabilities $P_j=|a_j|^2$  in  S measurement.
 In general
to calculate  $\Phi$ evolution for arbitrary complex system MS
 equations (\ref {CC}) 
for  MS Hamiltonian can be used and $P_j(t)$ found.
 Then from $P_j(t)$ at the time when S-$O$ interaction finished 
we find random $V^O$ which constitute stochastic
component of our  quantum dynamics. So if we have several S-$O$ rescatterings
each time after the interaction finished we get new $V^O$ state component
which effect in details will be discussed below.
 Note that practically we must be interested only in final $O$ state, because
in this model $O$ has no memory about intermediate  states.

We noticed already that in standard QM formalism in MS state (\ref {AA3})
$Q,Q_O$ aren't objectively existing for $O'$ which is serious argument against
Witnessing interpretation $\cite {Lah}$. That's true also in our
ansatz, but in its framework in the same time $Q$ and $Q_O$ have
objective values $Q_j, Q_{Oj}$ in $O$ RF. 
Note that the physical meaning of
Hilbert space $\cal H$ in our formalism 
 differs from standard
QM, because in it  all its Hermitian operators are $O$ observables, but
 the  operator $B$ of (\ref {AA8}) isn't observable for $O$.
Due to it we can speculate that $B$ observable for $O$ transformed into
stochastic parameter  $V^O_j$, which seems
 natural generalization of standard QM on observer evolution.
   Really $B$ describes observer $O$
internal state parameters, and so such transformation don't contradicts to
standard QM applicability for any external objects.
Realistic $O$ has many internal DFs  practically unobservable for him
and previously we assumed that they are responsible for 
randomness in the quantum  measurement \cite {May3}.

\section {Collapse and Quantum Memory Eraser}

To discuss measurement dynamics in our formalism for more
subtle situations let's consider several 
$gedankenexperiments$ for different
selfmeasurement effects, the first one means 
 'Undoing' the measurement. Such experiment was discussed by Vaidman
$\cite {Vaid}$  and Deutsch $\cite {De}$ for many worlds interpretation (MWI), 
 but we'll regard its slightly different version. Its first stage coincides
with regarded
S state (\ref {AAB}) measurement by $O$ 
 resulting in the final state (\ref {AA2}). 
This S measurement  can be undone
or reversed with the help of auxiliary devices - mirrors, etc.,
which come into action at $t>t_1$ 
and reflects $S$ back in $O$ direction and make them reinteract. 
It permit for the final state $\Psi_{MS}$ obtained at time $t_1$  at the
later time $t_2$ to be transformed backward to MS initial state $\Psi^0_{MS}$.
In any realistic layout to restore state (\ref {AAB}) is practically
impossible but to get the arbitrary S-$O$ factorized state by means of
such reversing is
more simple problem and that's enough for such tests.
Despite that under  realistic conditions  the decoherence processes 
make this reversing immensely difficult it doesn't contradict to
any physical laws.

 If we consider this experiment in standard QM 
from $O$ point of view  we come to  some nontrivial  conclusions. 
When memorization finished at $t_1$ in each event MS collapsed to some
 arbitrary state $|s_i\rangle|O_i\rangle$. Then at $t_2$  $O$ undergoes
the  external reversing influence, in particular it can be the
second collision with S during reversing experiment and its
state changes again and such
rescattering leads to a new state correlated with $|s_i\rangle$ :
$$
      |s_i\rangle|O_i\rangle \rightarrow |s'_i\rangle|O_0\rangle
$$
It means that $O$ memory erased and he
forgets Q value $q_i$, but if he measure S state again
he would restore the same $q_i$ value.
Its statistical state is
$$
   \rho'_m=|O_0\rangle \langle O_0| \sum |a_i|^2|s'_i\rangle \langle s'_i|
$$
 But this S final state differs from
MS state (\ref {AAB}) predicted from MS linear evolution observed by  $O'$ and
in principle this difference can be tested on S state  
without $O$ measurement. 
In our doublet formalism it's necessary also to describe subjective 
event-state component $V^O$ which
 after measurement  becomes some random $V^O_j$.
But after reversing  independently of $j$ it
returns to initial value $V^O_0$ , according to evolution ansatz ( \ref {CC})
 described in previous chapter.
If such description of this experiment is correct, as we can
believe because its results coincides with Schrodinger evolution in $O'$  
 it follows that after $q_i$ value erased from $O$ memory
 it lost unrestorably also
for any  other possible observer. If after that $O$ would measure Q again
 obtained by $O$ new value $q_j$ will have no correlation with $q_i$,
but $O$ can't make this comparison in principle.

Of course one should remember that  existing for finite time intermediate
$O$ states are in fact virtual states and differ from really stable 
states  used here, but for macroscopic time
 intervals this difference becomes
very small and probably can be neglected.

The analogy of 'undoing' with  quantum eraser experiment is straightforward :
there the photons polarization carry the information 
which can be erased and so change the system state $\cite {Scu}$.   
The analogous experiment with information memorization by 
some massive objects like molecules will be important test of
collapse models.

Note that observer $O'$ can perform on $O$ and S also the direct measurement
of interference terms for (\ref {AA2}) without reversing MS state.
Such experiment regarded for Coleman-Hepp model in  $\cite {May3}$
doesn't introduces any new features in comparison with 'Undoing'
and so we don't discuss it here.

The second experiment is the comparison of two independent measurements.
 In first one at the initial stage  
 $O$ measures $Q$ value of
 S at $t_1$ which results in
MS state (\ref {AA2}) for $O'$, but after it  $Q$ is measured again 
by observer $O'$ at $t_2>t_1$.
The interaction of $O'$ with MS results in entangled state of 
S,$O$ and $O'$  and so both observers acquire  some information 
about S state. This state vector in our formalism is:
\begin {equation}
   \Psi'_{MS}=|a_1 |s_1\rangle|O_1\rangle|O'_1\rangle+
a_2|s_2\rangle|O_2\rangle|O'_2\rangle
                                   \label {AAX}
\end {equation}   
   Such experiments  discussed  frequently due to
its relation to EPR-Bohm correlations and here we regard
 only its time  evolution aspects. Our question is : at what time
$Q$ value becomes certain and S state collapse occurs for $O'$ ? 
In our formalism at $t_1<t<t_2$ observer $O$ already acquired the information
that Q value is some  $q_i$, reflected by $\phi_I=|O_i\rangle\langle O_i|$.
 In the same time Q value stays
 uncertain  for $O'$, because relative to her MS state vector is (\ref {AA3})
,and $O'$ dual state is 
$$
     \Phi'=| \Psi_{MS} , |O'_0\rangle \langle O'_0| \gg
$$
 When at $t>t_2$  measurement by $O'$  
 finished  Q value measured by $O'$ 
coincides with $q_i$.
To check that Q value coincides for $O'$ and $O$, $O'$ can perform 
measurement both Q and $Q_O$ which is described by (\ref {AA22})
and gives the same result as in standard QM.  
It don't contradicts to the previous assumption 
 that for $O'$ before $t_2$ $Q$ was principally uncertain.
The reason is that in between $O'$ interacts with S and it
 makes Q value definite for him. 
This measurement   demonstrates the subjective character
of collapse, which happens only after  S interaction  with
particular observer occurs.
It contradicts with standard QM reduction postulate 
 which states that if Q acquired the definite value relative to 
$O$ then its objectively exists also for $O'$ or any other
observer. Then at $t>t_1$ MS state relative to $O'$ must be
the mixture $\rho_m$ of (\ref {AA3}). 
In our formalism at that time MS state vector relative to
$O'$  is pure state $\Psi_{MS}$ of (\ref {AA2})  which isn't $Q$
eigenstate.
 To test it experimentally $O'$ can measure $\hat{B}$ on MS
which don't commute with Q. If our theory is correct then 
$\bar{B} \ne 0 $ and thus MS state collapse doesn't occurs
at $t=t_1$.  
%Alternatevely  $O'$ can perform 'undoing' on MS and Q value
% known to $O$ will be erased unrestorably, which is impossible if
% Q value objectively existed for $O'$. 

It's worth to remind that the state vector has two aspects : dynamical
 and  informational in which $\Psi$ is $O$ maximal information
about the studied object S. Our formalism extends this aspect on
the case when $O$ measures S and can have more information about S
then 'stand-by' $O'$. In its framework the state collapse directly
related with $O$  information acquisition via interaction with S. 
 The same information can be send by  $O$ to another $O'$
by some material signal, for example photons bunch. When it measured by
$O'$ it result for her into  the entangled state $\Psi_{MS}$ collapse
to one of its components. Remind that in standard QM with reduction
 when $O$ measures S he also must send material signal to $O'$
which she must measure to acquier information on $Q$ value. So
despite the formal difference between two theories is large
the operational difference isn't so significant.

Relativistic analysis of EPR-Bohm pairs measurement  also indicates
subjective character of state vector and its collapse $\cite {Aha2}$.
 It was shown
that  the state vector can be defined only on space-like hypersurfaces
which are noncovariant for different observers.
This results correlate with nonequivalence of different observers
 in our  nonrelativistic formalism.
Hence we believe EPR-Bohm correlations
 deserve the detailed study in this dual framework. 

%\begin {equation}
%   \Psi_{MS} \rangle =a_j |s_j\rangle|O_j\rangle+
%\sum_{i\ne j}  a_j|s_i\rangle|O_i\rangle
                                   \label {BB2}

%To improve  the situation  the results of previous chapter can be used,
% then the  state (\ref {AA3}) corresponds to MS subjective state and
%to assume that MS  interaction even relative to $O$ described by 
% the complete MS state equivalent to (\ref {AAB}).

\section {Discussion}

In this paper the measurement models which accounts observer (IGUS) information
processing and memorization regarded. Real IGUSes are very complicated systems 
with many DFs, but the main quantum effects 
 are supposedly the same for large and small systems and 
can be studied with the simple models. 
 In our approach  observers are material local objects which are nonequivalent
in a sense that the  physical world description can be principally different
for each of them \cite {Rov}. 
Breuer theorem shows that inclusion of observer as quantum object
into measurement model doesn't explains collapse appearance. 
To obtain it it's neccessary also to modify the quantum states set
which makes it nonequivalent for different observers but conserves
Schrodinger dynamics for each state.

Our doublet formalism demonstrates that probabilistic represention
 is generic and unavoidable for QM and without it QM can't
acquire any observational realization. Wave-particle dualism
was always regarded as characteristic QM  feature, but in our formalism
it has straightforward  description. 
This dual theory in its present form
 is  the essencially phenomenological one. It don't 
answers: ' why state collapse exists in QM ?', but describes for which
quantum states set  it can be done  compatible with MS Schrodinger
evolution observed by external $O'$. In our opinion its results
evidences that the quantum state by its nature is closer to
classical probabilistic distribution then to De Brogile wave, despite 
it doesn't mean hidden parameters existence. To investigate its physical 
meaning we propose Information Causality Interpretation
which we'll be reported in forcoming paper \cite {May11}.
Here we note only that the key problem becomes  the
existence of the objective reality of any physical values i.e values
independent of particular observer. Our formalism hints that no
such reality is possible and any value has only subjective reality
relative to particular observer. 

Note that our formalism is principally different from Hidden Parameters
theories where this stochastic parameters influence Quantum state
dynamics and  so differs from SD.
 In our model  $V^O$ internal parameter $j$ is on the opposite
 controlled by evolution equation for quantum state, but don't exists
objectively before S starts to interact with $O$.

In dual formalism $O$ percepts only $O_j$ component of 
complete state vector $\Psi_{MS}$ 
and  it's not clear why other components aren't observed.
The tempting explanation can be related to  Breuer theorem result
 which shows that $O$ 
selfmeasurement is always noncomplete. It's possible to assume
that  in individual  event $O$ can percept only random part  $O_j$ of
his effective physical state. Thus $V^O(n)$ information loss
can be regarded as the consequence
stochastic degeneration of $\Psi_{MS}$ and to get more clear picture
one should consider statistical state $R_O$. But it also doesn't 
give complete MS description as was shown in Breuer papers $\cite {Bre}$.
% Note that in the same time $O'$ can perform $B$ measurement which
% demonstrates the presence of other $O$ components.
 In addition remind that
S initial state vector describes $Q$ fundamental uncertainty  for $O$ i.e.
its subjective information on S. When S-$O$ interacted
corresponding $O$ internal DFs excited and its internal state correlated  
with S. It's possible to assume that for $O$ internal states any
uncertainty is excluded - i.e. $O$ knows his own state
due to continuous interactions inside $O$ and initially uncertain $Q$
 percepted as random but certain value. $O$ is the 'last ring' of
measurement chain and such singularity can appear.

It's widely accepted now that decoherence effects are very important
in measurement dynamics $\cite {Zur,Gui}$. But the frequent claim
that collapse phenomena  can be completely explained in its framework
was shown to be incorrect at least for simple models $\cite {Desp}$.
But in our model some kind of decoherence
is also present in the form of self-decoherence when S departs from
$O$ volume after interaction.  In our model additional decoherence
effects must appear via account of the interaction of $O$ with its environment
    E and its effects can be importatnt.
Our approach to collapse is close to the decoherence attitude,
where also any additional collapse  postulate don't used. The main difference
is  that Decoherence theory claims the collapse is objective phenomena
which  was proved to be true only in some crude approximation
 \cite {Desp}. We suppose
that collapse has relational or subjective character and observed only by
observer inside decohering system, while for external $O$ this system  
including E is pure.

As we proposed in the introduction our theory doesn't need any
addressing to to human observer consciousness (OC). Rather in this model 
$O$ is active RF which internal state excited by the interaction
with the studied object. 
The situation with the measurement problem for two quantum observers
has much in common with Quantum reference frames introduced by
 Aharonov  $\cite {Aha3}$.

Historically the possible influence of observer on measurement
process was discussed first by London and Bauer \cite {Lon}. They supposed
that OC due to 'introspection
action' violates in fact Schrodinger equation for MS 
and results in state reduction. This idea was critisized in detail 
by Wigner \cite {Wig}. In distinction in our dual theory OC perception     
doesn't violate MS Schrodinger evolution from $O'$ point of view.
But measurement  subjective perception in it also performed by OC
and its results partly independent of dynamics due to
its dependence on stochastic $V^O$. This effect deserves further
discussion, but we believe that  such probabilistic behavior
is general IGUS property not related to OC only.

Dual formalism deserves detailed comparison with formalisms of different MWI 
variants, due to their analogy  - both are the theories without
dynamical collapse  $\cite {Busch}$.
In Everett+brain QM interpretations eq. ($\ref {AA3}$)
describes so called observer $O$ splitting identified with state collapse
 $\cite {Whi}$.  In this theory it's  assumed that each
$O$ branch describes the different reality and the state collapse is
phenomenological property of human consciousness. Obviously this
approach has
some common points with our models which deserve further analysis. 
In general all our experimental conclusions are based on human
subjective perception. Assuming the computer-brain perception analogy
in fact means that human signal perception also defined by $\bar{Q}_O$
values. Despite that this analogy looks quite reasonable we can't
give any proof of it.  In our model
in fact  the state collapse have subjective character and 
occurs initially only for single observer $O$ $\cite {Rov}$.
Copenahgen interepretation based heavily on the
micro/macro world partition. Our theory indicates that  
if it's sensible to discuss any world partition prompted by QM results
it seems to be the division between the subject  and  the objects.
Here subject is observer $O$
 which collect information
about surrounding world  and  the objects   can include other
observer $O'$.


\begin {thebibliography}{99}


\bibitem {Busch} P.Busch, P.Lahti, P.Mittelstaedt,
'Quantum Theory of Measurements' (Springer-Verlag, Berlin, 1996)

\bibitem {Gui} D.Guilini et al., 'Decoherence and Appearance of
Classical World', (Springer-Verlag,Berlin,1996) 

\bibitem {Alb} D.Z.Albert, Phyl. of Science 54, 577 (1986)
 
\bibitem {Pen} R.Penrose, 'Shadows of Mind' (Oxford, 1994) 

\bibitem {Wig} E.Wigner, 'Scientist speculates' , (Heinemann, London, 1962)

\bibitem {Rov}  C. Rovelli, Int. Journ. Theor. Phys. 35, 1637 (1995); 
quant-ph 9609002 (1996), 

\bibitem {Koch} S.Kochen 'Symposium on Foundations of Modern Physics'
, (World scientific, Singapour, 1985)
 
\bibitem {Bre} T.Breuer, Phyl. of Science 62, 197 (1995),
 Synthese 107, 1 (1996)

\bibitem {May3} S.Mayburov Quant-ph 9911105 


\bibitem {Elb} A.Elby, J.Bub Phys. Rev. A49, 4213, (1994)

\bibitem {Gir} GC. Girardi, A.Rimini, T.Weber Phys.Rev. D34, 470 (1986) 

\bibitem {Zur} W.Zurek, Phys Rev, D26,1862 (1982)

\bibitem {Lah} P. Lahti Int. J. Theor. Phys. 29, 339 (1990)

%\bibitem {Don} M.Donald, Found. Phys 25, 529 (1995)

\bibitem {Scu} M.Scully, K.Druhl Phys. Rev. A25, 2208 (1982) 


\bibitem {Shum} B. Schumaher , Phys. Rev. A51, 2738 (1995)

\bibitem {Vaid} L.Vaidman, quant-ph 9609006 

\bibitem {De} D.Deutsch, Int J. Theor. Phys. 24, 1 (1985)

\bibitem {Hep} K.Hepp, Helv. Phys. Acta 45 , 237 (1972)


%\bibitem {May5} S.Mayburov, Int. Journ. Theor. Phys. 34,1587(1995)





%\bibitem {Lan} R.Landauer , Phys. Lett. A217, 188 (1996)

\bibitem {Desp} W. D'Espagnat, Found Phys. 20,1157,(1990)

\bibitem {Sna} S.Snauder Found., Phys. 23, 1553 (1993) 

\bibitem {Nam} M.Namiki, S.Pascazio, Found. Phys. 22, 451 (1992)

\bibitem {Ume} H.Umezawa,H.Matsumoto, M.Tachiki, 'Thermofield 
Dynamics and Condensed States' (North-Holland,Amsterdam,1982)

\bibitem {Fuk} R. Fukuda, Phys. Rev. A ,35,8 (1987)

\bibitem {May2} S.Mayburov, Int. Journ. Theor. Phys. 37, 401 (1998)

\bibitem {Jan} Jansson B. 'Random Number Generators' (Stokholm, 1966)

\bibitem {Aha2} Y.Aharonov, D.Z. Albert Phys. Rev. D24, 359 (1981)

\bibitem {May11} S.Mayburov 'Information Causality Interpretataion
of Quantum Mechanics and Quantum Space-time' ,
 Talk given on II Quantum Gravity Workshop, Dubna, 2000
(to appear in proceedings)

\bibitem {Lon} London F., Bauer E. La theorie de l'Observation
(Hermann, Paris, 1939)   

\bibitem {Aha3} Y.Aharonov, T.Kaufherr Phys. Rev. D30, 368 (1984)
 
\bibitem {Whi} A.Whitaker, J. Phys., A18 , 253 (1985)


\bibitem {Bel} J.S.Bell, Helv. Phys. Acta 48, 93 (1975)

\end {thebibliography}
\end{document}